\documentclass[prd,twocolumn,showpacs,superscriptaddress,preprintnumbers,nofootinbib,
amsmath,amssymb]{revtex4-1}

\usepackage{graphicx} % Required for inserting images
\usepackage{tikz}
\usetikzlibrary{arrows}

\usepackage[colorlinks=true,urlcolor=blue,linkcolor=blue,citecolor=blue]{hyperref}
\usepackage[normalem]{ulem}
\usepackage[capitalize]{cleveref}

\usepackage[export]{adjustbox}
 
\usepackage{comment}
\usepackage{flushend}

\usepackage[T1]{fontenc} % if needed

\usepackage{mathtools}

\usepackage{xcolor}
\usepackage{graphicx}% Include figure files
\usepackage{dcolumn}% Align table columns on decimal point
\usepackage{bm}% bold math
\usepackage{multirow}

\usepackage{epstopdf}

\newcommand{\Beq}{\begin{equation}\begin{aligned}}
\newcommand{\Eeq}{\end{aligned}\end{equation}}

\begin{document}

\preprint{YITP-24-84}

\title{Crunch from AdS bubble collapse in unbounded potentials} 

\author{Kaloian D. Lozanov}
\email{kaloian.lozanov@ipmu.jp}
\affiliation{Kavli Institute for the Physics and Mathematics of the Universe (WPI), UTIAS
The University of Tokyo, Kashiwa, Chiba 277-8583, Japan.}
\author{Misao Sasaki}
\email{misao.sasaki@ipmu.jp}
\affiliation{Kavli Institute for the Physics and Mathematics of the Universe (WPI), UTIAS
The University of Tokyo, Kashiwa, Chiba 277-8583, Japan.}
\affiliation{Center for Gravitational Physics and Quantum Information, Yukawa Institute for Theoretical Physics,
Kyoto University, Kyoto 606-8502, Japan}
\affiliation{Leung Center for Cosmology and Particle Astrophysics, National Taiwan
University, Taipei 10617, Taiwan}

\date{\today}

\begin{abstract}
We consider a scalar field theory with a Minkowski false vacuum and an unbounded (or very deep) true vacuum. We show compelling evidence that an AdS bubble of vanishing total energy, embedded in asymptotically flat spacetime, generically undergoes a spherical collapse which leads to a space-like curvature singularity after the formation of trapped surfaces and apparent horizons. The crunch singularity, which is hided behind an apparent horizon, occurs before the true vacuum is reached, and the existence of a lower bound of the scalar field potential is not a necessary condition for its formation. 
\end{abstract}

\maketitle

\section{Introduction}

Scalar fields with zero-energy false vacua and deep true vacua are commonplace in high energy physics, e.g., Bosonic String theory and the String theory landscape \cite{Polchinski:1998rq,Agmon:2022thq}, possibly the Standard Model Higgs \cite{Elias-Miro:2011sqh,Gialamas:2022gxv}, etc. In the context of early-universe cosmology, transitions of a scalar field into the negative true vacuum could lead to rich phenomenology, e.g., expanding and/or collapsing Anti-de Sitter (AdS) domains, primordial black hole formation, constraining inflationary scenarios~\cite{Espinosa:2015qea,Jain:2019wxo,DeLuca:2022cus}, and dark matter production scenarios ~\cite{Espinosa:2017sgp}.

The evolution of a gravitating bubble with a negative energy-density core has been mainly studied analytically, employing the thin-wall approximation \cite{Weinberg:2012pjx,Coleman:1980aw,Abbott:1985kr,Espinosa:2015qea,DeLuca:2022cus}. In recent years, there have been numerical \cite{Hwang:2012pj,Strumia:2022kez,Chew:2023olq} and analytical \cite{Dong:2011gx,Kanno:2011vm,Kanno:2012zf,Espinosa:2020qtq,Espinosa:2021tgx} studies (based on analytic continuations from Euclidean Instanton and Bounce solutions), going beyond the thin-wall approximation, for potentials of specific forms. 
It has been established that the interior of the AdS bubble undergoes a gravitational collapse, leading to the formation of trapped surfaces, and a  space-like singularity behind them, known as the AdS crunch. 
Depending on the details of the bubble configuration and its embedding \cite{Espinosa:2015qea}, e.g., if the total (ADM) mass is positive, zero or negative, and on whether the asymptotic spacetime is dS, Minkowski or AdS, the exterior of the bubble could collapse -- leading to a black hole, or expand -- engulfing the entire false vacuum.

In this work we study for the first time the gravitational collapse of an AdS bubble of a scalar field with a Minkowski false vacuum and an unbounded (or very deep) true vacuum. 
We show that the existence of a lower bound of the scalar field potential is not a necessary condition for the formation of trapped surfaces and apparent horizons. 
We discover an apparent horizon can be generated while the scalar field is rolling down the slope of the unbounded potential (or towards its deep true vacuum, if the potential is bounded). Namely, a naked singularity does not form.
This is in contrast with the commonly accepted picture that trapped surfaces appear after the true vacuum is reached. 

We consider a natural scalar field potential -- a quadratic false vacuum, and an infinite (or very deep) true vacuum, approached along a quartic slope. We do not employ the thin wall approximation. We assume the bubble collapse follows its nucleation through an $O(4)$-symmetric instanton, known as the Coleman-De Luccia instanton. 
We also assume the subsequent bubble collapse is $O(3,1)$ symmetric, reducing its equations of motion to a system of ODEs which is readily solved. 

The validity of the $O(3,1)$ symmetry is straightforwardly proven for the bubble exterior. 
If the solution is analytic, it immediately follows that the bubble interior is also $O(3,1)$ symmetric.  
However, it needs to be shown in general, as the solution may not be analytic. 
To this end, we independently numerically solve the Einstein and Klein-Gordon PDEs, using different coordinate system and slicing condition, without assuming $O(3,1)$ symmetry but only spherical symmetry. 
We employ the 3+1 ADM formalism. The simulations confirm the $O(4)$ initial data gives rise to a globally $O(3,1)$ bubble, validating our ODE approach and its conclusions.

We observe the scalar field continuously speeds up its roll along the steepening negative slope, the positive kinetic energy grows, eventually overtakes the negative potential energy, rendering the core energy density positive. The positive energy-density core leads to the formation of trapped surfaces, and the subsequent rapid unbounded growth of the energy density (and associated geometric invariants) signifies the formation of the crunch singularity. This is in accordance with the notion that the formation of trapped surfaces is one of the sufficient (along with others) conditions for geodesic incompleteness, i.e., singularity existence \cite{Hawking:1973uf}.

The work is organised as follows. In Section \ref{sec:Model} we present the details of the model. Section \ref{sec:Bubnuc} outlines the initial conditions for the bubble collapse, which are based on the quantum bubble nucleation from a false vacuum. The subsequent bubble evolution is studied in Section \ref{sec:Bubevolve}. 
We present our summary in Section \ref{sec:Conclusions}. 
Technical details about scaling symmetries of the model, numerical GR studies and their results are delegated to the appendices.

Throughout the paper, we work in units in which $\hbar=c=1$, and the reduced Planck mass is $m_{pl}=1/\sqrt{8\pi G_N}$. We use the Einstein summation convention for repeated Greek and Latin indices.

\section{The model}
\label{sec:Model}

We consider a real, canonical scalar field, minimally coupled to Einstein gravity, governed by the action,
\Beq
S=\int d^4x\sqrt{-g}\left[\frac{m_{pl}^2}{2}R-\frac{(\partial\varphi)^2}{2}-V(\varphi)\right]\,,
\Eeq
with the potential,
\Beq
V(\varphi)=\left(1-\frac{\varphi^2}{M^2}\right)\frac{m^2\varphi^2}{2}\,.
\label{potential}
\Eeq
The scalar field stress-energy tensor is
\Beq
T_{\mu\nu}=\partial_\mu \varphi\partial_\nu \varphi -\frac{g_{\mu\nu}}{2}\left[\partial_\alpha\varphi\partial^\alpha\varphi+2V\right]\,,
\Eeq
which sources the Einstein equations,
\Beq
G_{\mu\nu}=\frac{T_{\mu\nu}}{m_{pl}^2}\,.
\Eeq
The classical evolution of $\varphi$ is governed by the Klein-Gordon equation,
\Beq
\Box \varphi-V'(\varphi)=0\,.
\Eeq

\begin{figure}[t]
\includegraphics[width=2.5in]{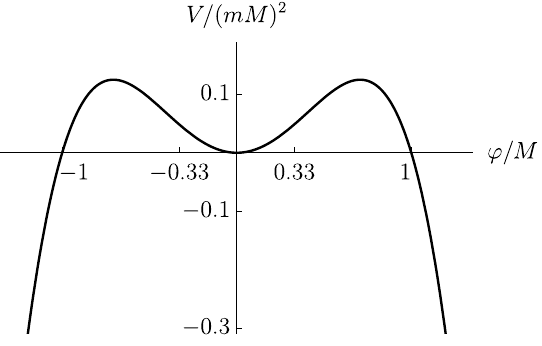}
\caption{False-vacuum potential with unbounded (or very deep) true vacuum.}
    \label{fig:Potential}
\end{figure}

The scalar-field potential has a local quadratic minimum (i.e., a false vacuum) at\footnote{For a related study in which $\varphi=0$ is a global maximum, considering Fubini instanons non-minimally coupled to gravity, see Ref. \cite{Tetradis:2023fnu}.} $\varphi=0$, becomes negative for $\varphi^2>M^2$, and is unbounded from below ($V(\varphi)\rightarrow-\infty$) when $\varphi\rightarrow\pm\infty$ (see Fig. \ref{fig:Potential}). 
Since $V(\varphi)$ can be negative, the scalar field violates the weak energy condition, and consequently, the weak cosmic censorship conjecture \cite{1969NCimR...1..252P}. 
This allows for the possibility of forming a naked singularity from generic non-singular initial data.

Before solving the system, we mention an important scaling property of the theory with the potential \eqref{potential}.
As explained in Appendix \ref{App:rescale}, 
the equations of motion and the solutions can be made to depend only on the parameter $\mu\equiv M/m_{pl}$ by rescaling the scalar field and the coordinates by $\varphi\to\tilde{\varphi}=\varphi/M$ and $x^\mu\to \tilde{x}^\mu=mx^\mu$.

\section{Bubble nucleation}
\label{sec:Bubnuc}

We assume the initial data (the initial conditions) for the classical evolution of the collapsing bubble are set by a quantum tunneling transition -- a Coleman-De Luccia bounce \cite{Coleman:1980aw}. In particular, at early times ($t<0$), the scalar field is assumed to be in the metastable phase at $\varphi=0$, of zero total energy. At $t=0$ a decay towards the stable phase (for concreteness we choose $\varphi>M$) takes place. The decay process is a quantum tunneling effect. The subsequent evolution ($t>0$) of the new-phase bubble ($\varphi>M$) inside the old phase (asymptotically $\varphi=0$) is described classically, and is the subject of the next section. In the rest of the current section we derive the $\varphi$ and $g_{\mu\nu}$ configuration of the bubble at formation. 

\subsection{Formation}

We consider a spherically-symmetric bubble nucleation scenario in which the bubble is created at $t=0$ due to a quantum tunneling from the false vacuum. We assume $\varphi_{\rm fv}(t\leq0,r)=0$ and $g_{\mu\nu,\rm fv}(t\leq0,r)=\eta_{\mu\nu}$, corresponding to the vanishing total (scalar and gravitational) energy. 
Since quantum tunneling conserves total energy, the initial and final states of the transition process at $t=0$ are of the zero energy. 

To find the initial data after bubble nucleation, 
$\varphi_{\rm b}(t=0,r)\neq0$ and $g_{\mu\nu,\rm b}(t=0,r)\neq\eta_{\mu\nu}$,
we first solve the equations in the Euclidean time under the assumption of $O(4)$ symmetry. Such a solution in Euclidean spacetime is called the Coleman-De Luccia instanton. 
We performed a Wick rotation into imaginary time, $t\rightarrow i\tau$ ($g_{\mu\nu,\rm fv}\rightarrow\delta_{\mu\nu}$), to obtain the Euclidean action,
\Beq
\label{eq:Eucl}
S_E=\int d^4x_E\sqrt{g_E}\left[-\frac{m_{pl}^2}{2}R_E+\frac{(\partial_E\varphi)^2}{2}-V(\varphi)\right].
\Eeq 
We take the form of the metric as
\Beq
\label{eq:EuclMetr}
ds^2_E=&g_{E,\mu\nu}dx_E^\mu dx_E^\nu=\omega^2(\xi)\left(d\xi^2+\xi^2d\Omega_3^2\right)\,,\\
&d\Omega_3^2=d\chi^2+\sin^2\chi (d\theta^2+\sin^2\theta d\phi^2)\,,
\Eeq
and assume that the scalar field is only a function of $\xi$, $\varphi=\varphi(\xi)$, in accordance with the $O(4)$ symmetry.
A nice feature of this metric form is that it explicitly captures the conformal flatness (i.e., vanishing of the Weyl tensor).
In terms of the Cartesian coordinates $(\tau,x^i)$, $\xi^2=\tau^2+r^2$ ($r^2=\delta_{ij}x^ix^j$, $i=1,\,2,\,3$).

It is commonly accepted that the $O(4)$ symmetric solution dominates the false vacuum decay, with the decay rate per unit volume given by (see, e.g., Ref. \cite{Weinberg:2012pjx}),
\Beq
\frac{\Gamma}{\mathcal{V}}=\mathcal{A}e^{-\mathcal{B}},
\Eeq
where $\mathcal{A}$ contains quantum loop corrections, and $\mathcal{B}=S_E[\varphi_{\rm b},g_{E,\mu\nu,\rm b}]-S_E[\varphi_{\rm fv},g_{E,\mu\nu,\rm fv}]$, where $S_E[\varphi_{\rm b},g_{E,\mu\nu,\rm b}]$ is the Euclidean action of the $O(4)$ symmetric solution and $S_E[\varphi_{\rm fv},g_{E,\mu\nu,\rm fv}]$ is that of the false vacuum. Although there is no rigorous proof, we assume this is the case. 

The Euclidean version of the Klein-Gordon equation, $\Box_E \varphi+V'(\varphi)=0$, yields
\Beq
\label{eq:EuclKG}
\varphi''(\xi)+\left(\frac{3}{\xi}+2\frac{\omega'(\xi)}{\omega(\xi)}\right)\varphi'(\xi)-\omega^2(\xi)\frac{dV}{d\varphi}=0\,.
\Eeq
The $\xi$-$\xi$ component of the Euclidean Einstein equations, $G_{E,\xi\xi}=T_{E,\xi\xi}/m_{pl}^2$, yields
\Beq
\label{eq:EuclEins00}
\frac{2}{\xi}\frac{\omega'(\xi)}{\omega(\xi)}+\left(\frac{\omega'(\xi)}{\omega(\xi)}\right)^2=\frac{\omega^2(\xi)}{3m_{pl}^2}\left[\frac{1}{2}\left(\frac{\varphi'(\xi)}{\omega(\xi)}\right)^2-V\right]\,.
\Eeq
Eqs. \eqref{eq:EuclKG} and \eqref{eq:EuclEins00} govern the Euclidean evolution of $\varphi(\xi)$ and the metric, $\omega(\xi)$. 

It is also useful to have the second order differential equation for $\omega(\xi)$, which may be obtained from the other components of the Einstein equations, or combining eqs. \eqref{eq:EuclKG} and \eqref{eq:EuclEins00}.
We obtain
\Beq
\label{eq:Euclddomega}
\frac{\omega''(\xi)}{\omega(\xi)}-\left(\frac{1}{\xi}+2\frac{\omega'(\xi)}{\omega(\xi)}\right)\frac{\omega'(\xi)}{\omega(\xi)}=-\frac{1}{2}\left(\frac{\varphi'(\xi)}{m_{pl}}\right)^2\,.
\Eeq
For the initial data that satisfy eq. \eqref{eq:EuclEins00}, we may use this equation instead of eq. \eqref{eq:EuclEins00}, i.e.,  eqs. \eqref{eq:EuclKG} and \eqref{eq:Euclddomega}, instead of eqs. \eqref{eq:EuclKG} and \eqref{eq:EuclEins00},
when solving for $\varphi(\xi)$ and $\omega(\xi)$.

\begin{figure}[t]
\includegraphics[height=3.4in]{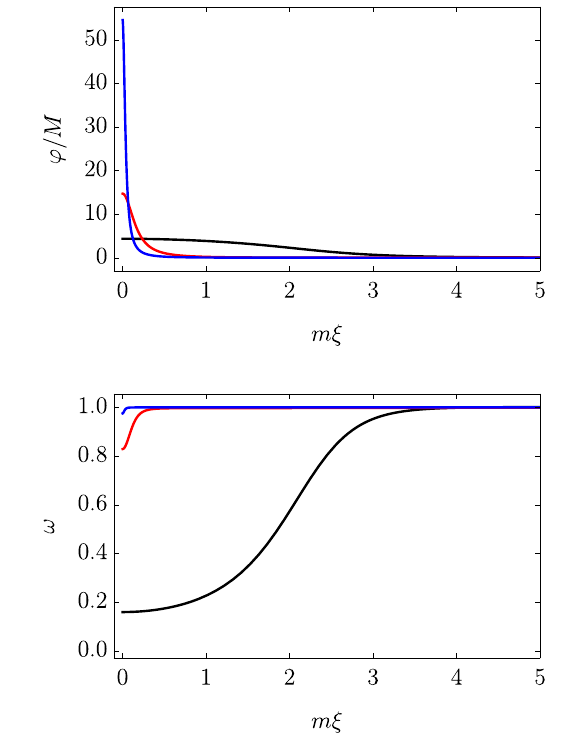}
\caption{Coleman-De Luccia bounce solution for the model parameter $\mu=M/m_{pl}=$ $1$ (black), $10^{-1}$ (red), $10^{-2}$ (blue). The solution can be also mapped on the future field and metric data of the subsequent evolution after nucleation, describing the bubble exterior, see Fig. \ref{fig:LiCoInterPhi}.}
    \label{fig:Euclphixi}
\end{figure}

The Coleman-De Luccia solution is monotonous and obeys the boundary conditions; $\varphi'|_{\xi=0}=0$, $\varphi|_{\xi\rightarrow\infty}=0$, $\omega'|_{\xi=0}=0$ and $\omega|_{\xi\rightarrow\infty}=1$. 
To find it, we use these boundary conditions to solve eqs. \eqref{eq:EuclKG} and \eqref{eq:Euclddomega}, according to the shooting method.\footnote{We solve eqs. \eqref{eq:EuclKG} and \eqref{eq:Euclddomega} from an infinitesimal $\xi=\xi_{\rm min}=\epsilon/m$ ($0<\epsilon\ll1$) until $\xi=\xi_{\rm max}\gg m^{-1}$ for various $\varphi_0$. We put $\varphi(\xi_{\rm min})=\varphi_0+\xi_{\rm min}\varphi'(\xi_{\rm min})/2$, $\varphi'(\xi_{\rm min})=\xi_{\rm min}\omega_0^2V'(\varphi_0)/4$, $\omega(\xi_{\rm min})=\omega_0+\xi_{\rm min}\omega'(\xi_{\rm min})/2$, $\omega'(\xi_{\rm min})=-\xi_{\rm min}\omega_0^3V(\varphi_0)/(6m_{pl}^2)$, and $\omega_0=1$. We vary $\varphi_0$ until $\varphi(\xi_{\rm max})/M=\varphi'(\xi_{\rm max})/(mM)=\mathcal{O}(10^{-5})$, which also implies $\omega'(\xi_{\rm max})/m\ll1$. We then re-scale $\omega\rightarrow \kappa \omega$ and $\xi\rightarrow \kappa^{-1}\xi$, which is a symmetry transformation, to make $\omega({\xi_{\rm max}})=1$. Typically, $\kappa=\mathcal{O}(1)$.}
In Fig. \ref{fig:Euclphixi} we present the Coleman-De Luccia solution for various model parameters. 

As noted before, by introducing the dimensionless scalar field $\tilde{\varphi}=\varphi/M$ and the dimensionless spacetime coordinates $\tilde{x^\mu}=mx^\mu$, there remains only one free dimensionless parameter in the problem,
\Beq
\mu\equiv \frac{M}{m_{pl}}\,.
\Eeq
We also note that as $\mu$ is reduced from unity, the gravitational effects become less pronounced, e.g., $\omega$ deviates less from $1$, and $\varphi$ becomes more localized.
%-- since it is gradient, as opposed to gravitationally supported.

\section{Bubble evolution}
\label{sec:Bubevolve}

To understand the gravitational bubble evolution for $t\geq0$ it is instructive to begin with the Lorentzian signature spacetime metric. In particular, after a Wick rotation, $\tau\rightarrow -it$, we have
\Beq
\xi^2=-t^2+r^2\,,
\Eeq
and the conformal flat Euclidean metric \eqref{eq:EuclMetr} becomes
\Beq
\label{eq:Lorsignmetric}
ds^2=\omega^2(\xi)(-dt^2+dr^2+r^2d\Omega_2^2)\,.
\Eeq
This metric covers the entire future spacetime and holds for both $\xi^2\geq0$ (i.e., on and outside the light cone, where $r\geq t$) and $\xi^2<0$ (i.e., inside the light cone, where $r<t$). 

We note that if we make an analytic continuation to the Lorentzian signature directly from the coordinates ($\xi,\chi,\theta,\phi$), it is necessary to make the following two separate analytic continuations (one for each region) of the angular variable $\chi$. Namely,
\begin{align}
&\chi=\frac{\pi}{2}+i\tilde{\chi}\quad&{\rm for}~\xi^2\geq0\,,
\label{eq:chiext}\\
&\chi=i\tilde{\chi} &{\rm for}~\xi^2<0\,.
\label{eq:chiint}
\end{align}
The results are related to the ($t,r$) coordinates, respectively, as
\begin{equation}
\begin{cases}
  t&=\xi\sinh\tilde{\chi}\\
  r&=\xi\cosh\tilde{\chi} 
\end{cases}
\quad {\rm for}~\xi^2\geq0\,,
\end{equation}
and
\begin{equation}
\begin{cases}
  t&=\tilde\xi\cosh\tilde{\chi}\\
  r&=\tilde\xi\sinh\tilde{\chi} 
\end{cases}
\quad {\rm for}~\xi^2<0\,,
\end{equation}
where we have set $\xi=i\tilde{\xi}$.

We also note that since $\xi^2\geq0$ on the initial data slice, and on and outside the light cone, the Euclidean solution (which is independent of $\chi$) also describes the Lorentzian-signature solution for $r\geq t$. On the other hand, the solution in the interior of the light cone ($\xi^2<0$) can be obtained either numerically or by an analytic continuation (if possible) to imaginary $\xi$ starting from the light cone at $\xi^2=0$. 
See Fig. \ref{fig:LiCoInterPhi} for a schematic representation of the geometry of the problem.

In the remainder of this section we consider in detail the evolution of the bubble in the two separate regions, namely the exterior and the interior of the future light cone.

\subsection{Light cone and its exterior}

On the light cone and in its exterior, the square of the coordinate $\xi$ is non-negative ($\xi^2\geq0$), just like in Euclidean space. Hence, the $O(4)$-symmetric Coleman-De Luccia solution from Fig. \ref{fig:Euclphixi} also describes the $O(3,1)$-symmetric solution in the domain $r\geq t$, i.e., $\varphi(t,r)=\varphi(\sqrt{-t^2+r^2})=\varphi(\xi)$ and $\omega(t,r)=\omega(\sqrt{-t^2+r^2})=\omega(\xi)$. The iso-$\varphi$ and iso-$\omega$ surfaces are hyperbolas of constant $\xi^2>0$ (and the light-cone line, $\xi^2=0$) along which $\tilde{\chi}$ from eq. \eqref{eq:chiext} grows from $0$ (at $t=0$) to $\infty$, see Fig. \ref{fig:LiCoInterPhi}. The constant values of $\varphi$ and $\omega$ along the hyperbolas and the light cone are given by the Euclidean solution for the corresponding $\xi$, see Fig. \ref{fig:Euclphixi}.

\subsection{Light-cone interior}

\begin{figure}[t]
\includegraphics[height=3.25in]{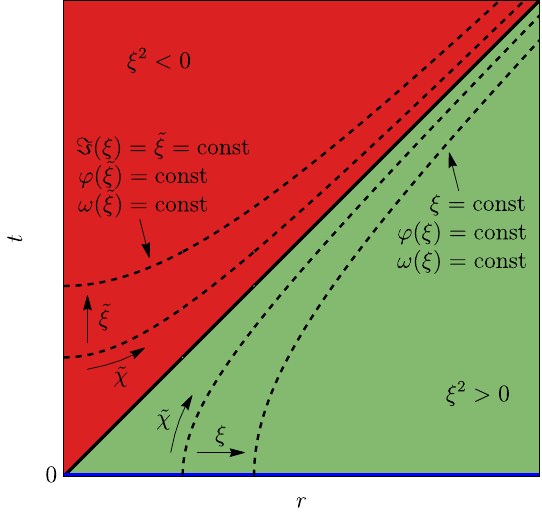}
\caption{A schematic representation, after the initial $O(4)$ Coleman-De Luccia transition (blue line at $t=0$), of the evolution of the AdS bubble in the $O(3,1)$-symmetric spacetime metric from eq. \eqref{eq:Lorsignmetric}. The light-cone line, along which $\xi=\sqrt{-t^2+r^2}$ is $0$, is given by the black solid line. In the light-cone exterior ($r>t$, green shaded area), worldlines of $\xi={\rm const}.$ are depicted by the black dashed hyperbolas, along each of which the hyperbolic angle coordinate $\tilde{\chi}$ from eq. \eqref{eq:chiext} grows from $0$ on the $r$ axis. Similarly, in the light-cone interior ($r<t$, red shaded area) we have black dashed hyperbolic slices of constant $\tilde{\xi}=\Im(\xi)$, along each of which $\tilde{\chi}$ from eq. \eqref{eq:chiint} grows from $0$ on the $t$ axis. For $r\geq t$, $\varphi(\xi)$ and $\omega(\xi)$ are constant along the light-cone line and the exterior hyperbolic worldlines, and are given directly by the Colemann-De Luccia data, e.g., see Fig. \ref{fig:Euclphixi}. For $r< t$, $\varphi(\tilde{\xi})$ and $\omega(\tilde{\xi})$ need to be calculated after imposing boundary conditions on the light cone (e.g., see Fig. \ref{fig:LiCoInterphi}), or by analytic continuation of the exterior solution. Finally, given the initial $O(4)$ data, the assumption of an $O(3,1)$ symmetry is robust for the exterior, but not the interior. We confirm the latter also respects the symmetry, by running ADM $3+1$ simulations of the bubble collapse which do not use $O(3,1)$ spacetime slicing. }
    \label{fig:LiCoInterPhi}
\end{figure}

The whole initial data slice has real $\xi$ ($\xi^2\geq0$) and is mapped onto the light cone and its exterior. The initial $O(4)$ Euclidean symmetry is inherited as $O(3,1)$ symmetry by the light-cone exterior, see Fig. \ref{fig:LiCoInterPhi}.

On the other hand, the interior of the light cone has imaginary $\xi$ ($\xi^2<0$) and can be related to the initial Euclidean solution only through an analytic continuation from real to imaginary $\xi$. Like the exterior, the interior is also $O(3,1)$ symmetric. We should note that two implicit assumptions are involved here: (i) the $\varphi(\xi)$ and $\omega(\xi)$ are analytic across and within the light cone (implying $O(3,1)$ symmetry of the light-cone interior), and (ii) the light-cone interior is $O(3,1)$ symmetric (which follows from, but does not imply analyticity).

Assuming analyticity (and the implied $O(3,1)$ symmetry), we can determine the bubble evolution (of $\varphi$ and $\omega$) within the light cone, $r<t$, by solving a simple system of ODEs, akin to eqs. (\ref{eq:EuclKG}-\ref{eq:Euclddomega}) with appropriate Wick-rotation sign corrections, as we show below. To check the validity of the assumptions, we also independently solve, with full numerical GR simulations based on ADM formalism for the Einstein PDEs, for the bubble evolution in the entire spatial domain (both outside and inside the light cone) in a different, generic coordinate system, which only assumes spherical symmetry, but no analyticity (and the implied $O(3,1)$ symmetry). The simulations results are in agreement with the solutions of the ODEs, confirming that the light-cone interior solution can be treated analytically (and is $O(3,1)$ symmetric), see Appendix \ref{App:bubble3plus1} and Fig. \ref{fig:rhosim}.

\begin{figure}[t]
\includegraphics[height=3.4in]{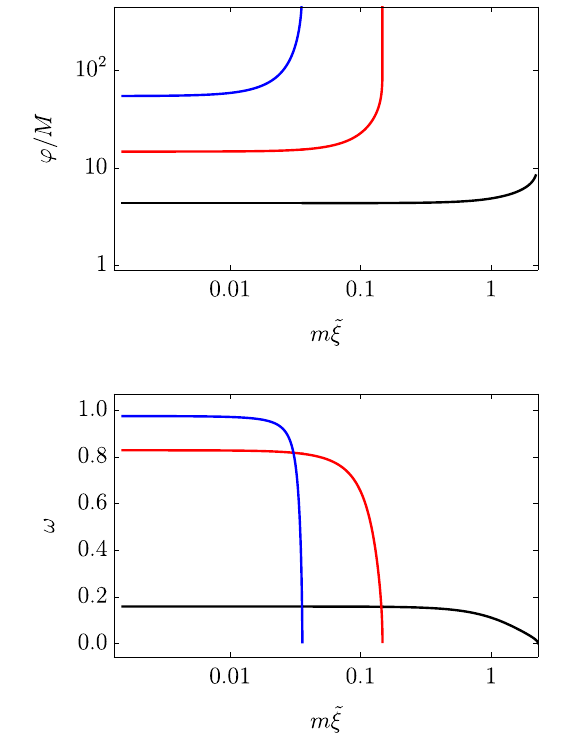}
\caption{Same as Fig. \ref{fig:Euclphixi}, but for the $O(3,1)$-symmetric light-cone interior, $\xi^2<0$, where $\tilde{\xi}=\Im(\xi)$ (see Fig. \ref{fig:LiCoInterPhi}). We use the analytic boundary conditions from eq. \eqref{eq:ICsInt} on the light cone $\tilde{\xi}\rightarrow0_+$.}
    \label{fig:LiCoInterphi}
\end{figure}

Within the light cone ($\xi^2<0$) of the $O(3,1)$-symmetric spacetime \eqref{eq:Lorsignmetric}, setting $\xi=i\tilde\xi$,
the Klein-Gordon equation reduces to
\Beq
\label{eq:LorLCintKG}
\varphi''(\tilde{\xi})+\left(\frac{3}{\tilde{\xi}}+2\frac{\omega'(\tilde{\xi})}{\omega(\tilde{\xi})}\right)\varphi'(\tilde{\xi})+\omega^2(\tilde{\xi})\frac{dV}{d\varphi}=0\,.
\Eeq
Similarly, the $\tilde\xi-\tilde\xi$ component of the Einstein equations yields
\Beq
\label{eq:LorLCEins00}
\frac{2}{\tilde{\xi}}\frac{\omega'(\tilde{\xi})}{\omega(\tilde{\xi})}+\left(\frac{\omega'(\tilde{\xi})}{\omega(\tilde{\xi})}\right)^2=\frac{\omega^2(\tilde{\xi})}{3m_{pl}^2}\left[\frac{1}{2}\left(\frac{\varphi'(\tilde{\xi})}{\omega(\tilde{\xi})}\right)^2+V\right]\,.
\Eeq
From eqs. \eqref{eq:LorLCintKG} and \eqref{eq:LorLCEins00}, we derive
\Beq
\label{eq:LorLCintddomega}
\frac{\omega''(\tilde{\xi})}{\omega(\tilde{\xi})}-\left(\frac{1}{\tilde{\xi}}+2\frac{\omega'(\tilde{\xi})}{\omega(\tilde{\xi})}\right)\frac{\omega'(\tilde{\xi})}{\omega(\tilde{\xi})}=-\frac{1}{2}\left(\frac{\varphi'(\tilde{\xi})}{m_{pl}}\right)^2\,.
\Eeq
We find the bubble evolution in the light-cone interior by solving eqs. \eqref{eq:LorLCintKG} and \eqref{eq:LorLCintddomega}.

Numerically, starting infinitesimally close to the light cone at $\tilde{\xi}=\tilde{\xi}_{\rm{min}}\rightarrow0_+$, we use initial conditions which assume analytic (series expansion) solutions near the light-cone line and satisfy eq. \eqref{eq:LorLCEins00} in this limit,
\Beq
\label{eq:ICsInt}
\varphi(\tilde{\xi}_{\rm min})&=\varphi_0-\frac{\tilde{\xi}^2_{\rm min}\omega_0^2V'(\varphi_0)}{8},\\  \omega(\tilde{\xi}_{\rm min})&=\omega_0+\frac{\tilde{\xi}^2_{\rm min}\omega_0^3V(\varphi_0)}{12m_{pl}^2}\,,
\Eeq
where the values of the constants $\varphi_0$ and $\omega_0$ are given by the Coleman-De Luccia solution at the origin (and on the light-cone line at $\xi=0$). The derivatives vanish by continuity.

\begin{figure}[t]
\includegraphics[height=3.4in]{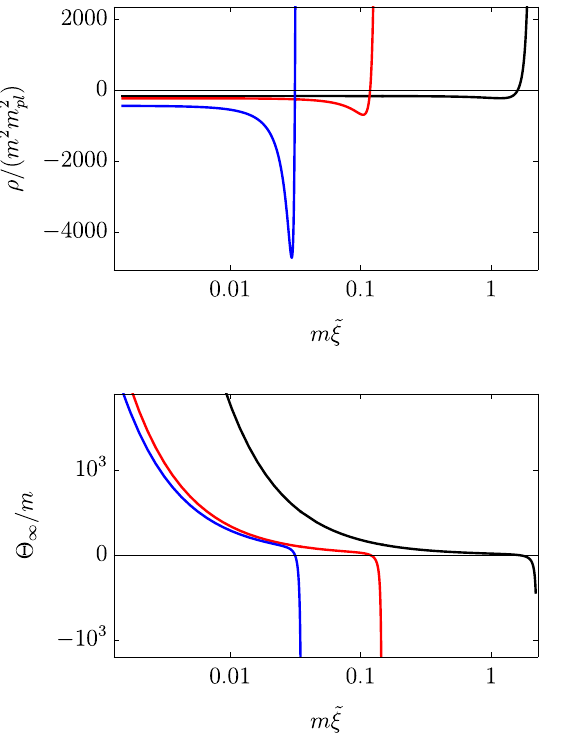}
\caption{The energy density, $\rho=\varphi'^2(\tilde{\xi})/(2\omega^2(\tilde{\xi}))+V$ (top panel), and the expansion rate of a null congruence, $\Theta_\infty$, at infinity, $\tilde{\chi}\rightarrow\infty$ (bottom panel), for the solutions from Fig. \ref{fig:LiCoInterphi}. Initially, while the scalar field rolls slowly down the slope of the negative potential, $V$, the energy density of the AdS bubble becomes increasingly negative. The growing positive kinetic energy at some point halts the change in $\rho$, and shortly afterwards dominates over $V$, making $\rho$ abruptly change from negative to positive. While $\rho$ is negative, no trapped surfaces and associate apparent horizon form, $\Theta_\infty>0$. The moment $\rho=0$, $\Theta_\infty$ also crosses zero (according to eq. \eqref{eq:ThRho}), signifying the formation of a trapped surface and apparent horizon at infinity. Subsequently, the positive $\rho$ (like related geometric invariants, e.g., the Kretschmann scalar given in Fig. \ref{fig:Kret}) quickly becomes infinite, which is a manifestation of a crunch-singularity formation.}
    \label{fig:LiCoInterrho}
\end{figure}

As the evolution proceeds within the light cone ($\tilde{\xi}$ increases from $0$ near the light cone, sweeping upwards the $t>r$ domain (see Fig. \ref{fig:LiCoInterPhi}) the scalar field $\varphi(\tilde{\xi})$ grows monotonically from the initial $\varphi_0$ to $\infty$ for a finite interval of $\tilde{\xi}$ (see the upper panel in Fig. \ref{fig:LiCoInterphi}), signifying the formation of a singularity. It is a crunch singularity, since the slices contract, i.e., the metric scale factor $\omega(\tilde{\xi})$ decreases monotonically from $\omega_0$ to $0$ (see the lower panel in Fig. \ref{fig:LiCoInterphi}).
However, it is not an AdS singularity, since the energy density, $\rho(\tilde{\xi})$, becomes positive before the singularity forms, as shown in Fig. \ref{fig:LiCoInterrho}. The evolution of the energy density is non-monotonic -- it starts out negative near the light cone, briefly decreases with $\tilde{\xi}$ to become more negative, until it reaches a minimum, when the scalar field speeds up its roll, its positive kinetic energy begins to overtake the negative potential energy, $V$, and $\rho$ rapidly increases to become positive and infinite. As we show next, the singularity is also not naked, i.e., it lies behind a series of trapped surfaces, with the outermost being the apparent horizon. Heuristically, this is expected, since the energy density on the singular $\tilde{\xi}$ slice is positive, whereas only negative energy densities (or masses) could yield naked singularities.

\subsubsection{Apparent-horizon formation}
\label{eq:Ahf}

Given that we are approaching a spacetime region of infinite energy density, we check if any trapped surfaces (and consequently any apparent horizons) form. To this end we implement a standard procedure.  In particular, we calculate the expansion of the outgoing null congruence. If the expansion ever becomes negative, we have a trapped surface, and the outermost such surface is the apparent horizon. 

To look for apparent horizons we use a standard construction, see, e.g., \cite{Baumgarte:2010ndz}. We embed a closed 2-dimensional hypersurface $\sigma$ in the 3-dimensional hyperbolic hypersurface of constant $\tilde{\xi}$, called $\Sigma$. The unit normal to $\Sigma$ is $n_\mu=(-\omega(\tilde{\xi}),0,0,0)$. $s^\mu$ is the outward unit normal to $\sigma$, obeying $s^\mu s_\mu=1$ and $s^\mu n_\mu=0$. 
The induced metric on $\Sigma$ is $\gamma_{\mu\nu}dx^\mu dx^\nu=\omega^2(\tilde{\xi})\tilde{\xi}^2(d\tilde{\chi}^2+\sinh^2\tilde{\chi}d\Omega_2^2)$, 
and the induced metric on the 2-sphere $\sigma$ is $m^{\mu\nu}=\gamma^{\mu\nu}-s^{\mu}s^{\nu}$.

We construct on $\sigma$ a pair of future pointing null-vector fields $k^\mu=(n^\mu+s^\mu)/\sqrt{2}$, $l^\mu=(n^\mu-s^\mu)/\sqrt{2}$. 
They are orthogonal to $\sigma$, but not $\Sigma$. They satisfy $m_{\mu\nu}k^\nu=0=m_{\mu\nu}l^\nu$, $k_{\nu}k^\nu=0=l_{\nu}l^\nu$, $k_{\nu}l^\nu=-1$, and $m^{\mu\nu}=\gamma^{\mu\nu}+k^{\mu}l^{\nu}+l^{\mu}k^{\nu}$. $k$ and $l$ are tangents, respectively, to a congruence of radially outgoing and ingoing null curves.

The expansion of the outgoing null congruence is defined as

\Beq
\Theta \equiv m^{\mu\nu} \nabla_\mu k_\nu\,.
\Eeq
An outer-trapped surface is a $\sigma$ on which $\Theta<0$ everywhere \cite{Hawking:1973uf}. A trapped region in $\Sigma$ contains outer-trapped surfaces. The outer boundary of a trapped region is a marginally trapped surface, defined by $\Theta_{AH}=0$ everywhere, and is also known as an (outer) apparent horizon.

\begin{figure}[t]
\includegraphics[width=2.8in]{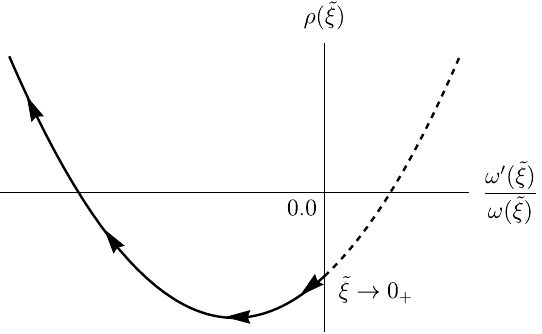}
\caption{Qualitative evolution (black solid line) of the energy density in the light-cone interior, $\rho=\varphi'^2(\tilde{\xi})/(2\omega^2(\tilde{\xi}))+V$, according to the Hamiltonian constraint, eq. \eqref{eq:LorLCEins00} (including black dashed line for $\omega'(\tilde{\xi})>0$, inapplicable to collapse, $\omega'(\tilde{\xi})<0$). This picture holds for every $V$ for which the crunch is approached at some point at a rate $(\omega(\tilde{\xi})\tilde{\xi}^2)'\leq0$, with equality corresponding to the zero crossing of the black solid $\rho$ line. Since this rate of contraction is attained during the collapse for a generic $V$, the occurrence of positive energy density, trapped surfaces and crunch singularities is expected for generic unbounded potentials.}
    \label{fig:LiCoInterrhosch}
\end{figure}

The expansion of the outgoing null congruence, for the spherical surface $\sigma$ centered at the origin with $s_\mu=(0,\omega(\tilde{\xi})\tilde{\xi},0,0)$, is given by
\Beq
\Theta(\tilde{\xi},\tilde{\chi})= \frac{\sqrt{2}}{\omega(\tilde{\xi})\tilde{\xi}}\left(\tilde{\xi}\frac{\omega'(\tilde{\xi})}{\omega(\tilde{\xi})}+1+\coth\tilde{\chi}\right)\,.
\Eeq
To assess if the singularity is globally naked, we consider the expansion rate at spatial infinity, namely,
\Beq
\label{eq:Thetinf}
\Theta_\infty(\tilde{\xi})\equiv\lim_{\tilde{\chi}\rightarrow\infty}\Theta(\tilde{\xi},\tilde{\chi})= \frac{\sqrt{2}}{\omega(\tilde{\xi})\tilde{\xi}}\left(\tilde{\xi}\frac{\omega'(\tilde{\xi})}{\omega(\tilde{\xi})}+2\right)\,.
\Eeq

Comparing the above expression with eq. \eqref{eq:LorLCEins00}, we find the exact relation, 
\Beq
\label{eq:ThRho}
\Theta_\infty(\tilde{\xi})=\frac{\sqrt{2}}{3}\frac{\omega^2(\tilde{\xi})}{\omega'(\tilde{\xi})}\frac{\rho(\tilde{\xi})}{m_{pl}^2}\,.
\Eeq
Since $\omega'(\tilde{\xi})<0$ always, and $\rho<0$ initially, $\Theta_\infty>0$ at the early stage.
As the system evolves, $\rho$ starts to increase, crosses zero, and becomes positive.
Here it is important to note that $\rho=0$ cannot be a reflection point. Namely, $\rho$ inevitably becomes positive once $\rho=0$ is attained. This can be seen by taking the derivative of $\tilde{\xi}\omega'(\tilde{\xi})/\omega(\tilde{\xi})$ at $\rho=0$ and evaluating it by using \eqref{eq:LorLCEins00} and \eqref{eq:LorLCintddomega}. It is negative definite at $\rho=0$, hence the decrease in $\omega'(\tilde{\xi})/\omega(\tilde{\xi})$ never stops at $\rho=0$.
Thus $\Theta_\infty(\tilde{\xi})$ changes its sign at $\rho=0$ from positive to negative, implying that an apparent horizon appears at infinity. 

As $\rho$ increases, $\omega'(\tilde{\xi})/\omega(\tilde{\xi})$ decreases further, as can be seen in Fig. \ref{fig:LiCoInterrho}.
Consequently, the locus of $\Theta(\tilde{\xi},\tilde{\chi})=0$ on the ($\tilde{\xi},\tilde{\chi}$) plane starts to move inward, given by $\coth\tilde\chi=-1-\tilde{\xi}\omega'(\tilde{\xi})/\omega(\tilde{\xi})$.
%In general, according to the Hamiltonian constraint \eqref{eq:LorLCEins00}, provided at some point $(\omega(\tilde{\xi})\tilde{\xi}^2)'<0$, $\rho$ changes from negative to positive, and we have the evolution depicted in Fig. \ref{fig:LiCoInterrhosch}.
%During the early stages of the bubble evolution (while $\rho(\tilde{\xi})<0$), $\Theta_\infty(\tilde{\xi})>0$ and there are no trapped surfaces. However, at the moment $\rho(\tilde{\xi})=0$, an apparent horizon forms at infinity, $\Theta_\infty(\tilde{\xi})=0$. 
At later times (larger $\tilde{\xi}$), trapped surfaces form at progressively smaller $\tilde{\chi}$ until the origin is reached ($\tilde{\chi}=0$), coinciding with the singularity formation.

Finally, to show that we have a genuine singularity, which should give rise to singular behaviour in geometric invariants, we consider the Kretschmann scalar,
\Beq
\tilde{\mathcal{K}}\equiv R^{\mu\nu\alpha\beta}R_{\mu\nu\alpha\beta}\,,
\Eeq
which inside the light cone takes the form,
\Beq
\tilde{\mathcal{K}}(\tilde{\xi}(t,r))=12\Bigg[&5\left(\frac{\omega'(\tilde{\xi})}{\tilde{\xi}\omega^3(\tilde{\xi})}\right)^2+\frac{2}{\tilde{\xi}\omega(\tilde{\xi})}\left(\frac{\omega'(\tilde{\xi})}{\omega^2(\tilde{\xi})}\right)^3\\
&+2\left(\frac{\omega'(\tilde{\xi})}{\omega^2(\tilde{\xi})}\right)^4+\frac{2\omega''(\tilde{\xi})\omega'(\tilde{\xi})}{\tilde{\xi}\omega^6(\tilde{\xi})}\\
&-\frac{2\omega''(\tilde{\xi})\omega'^2(\tilde{\xi})}{\omega^7(\tilde{\xi})}+\frac{\omega''^2(\tilde{\xi})}{\omega^6(\tilde{\xi})}\Bigg]\,.
\Eeq
One has an identical expression outside the lightcone, provided we make the replacement $\tilde{\xi}\rightarrow\xi$. 
As shown in Fig. \ref{fig:Kret}, $\tilde{\mathcal{K}}$ is constant along the slices of constant $t^2-r^2$, and grows exponentially fast, reflecting the formation of a singularity. 

\section{Conclusions}
\label{sec:Conclusions}

We studied the classical evolution of a spherically symmetric AdS bubble after nucleation from a Minkowski false vacuum, in a scalar field potential of an infinitely deep true vacuum, approached along a quartic slope. 
We observed the collapse of the interior of the AdS bubble and the formation of a space-like curvature singularity. We also saw the formation of trapped surfaces from the moment the energy density became positive. 
This shows that the formation of the AdS crunch behind an apparent horizon during the collapse of an AdS bubble does not require the existence of a lower bound of the scalar field potential. 
Our finding can be generalized to arbitrary potentials, except perhaps for extremely fine-tuned profiles. 
It may have interesting implications for models within the String Theory landscape paradigm. 

\acknowledgments
We would like to thank Sugumi Kanno and Jiro Soda for valuable comments on the draft. We are grateful to Nikolaos Tetradis for pointing out an incorrect comment about Fubini instantons non-minimally coupled to gravity in the previous version of the manuscript. This work is supported in part by JSPS KAKENHI Nos.~20H05853 and 24K00624.

\appendix

\section{Rescaling of the scalar field}
\label{App:rescale}
Here we show why there is only one free parameter in the theory, namely $\mu=M/m_{pl}$, although the potential is characterized by the two dimension-full parameters $m$ and $M$.

\begin{figure}[t]
\vspace{-0.3cm}
\includegraphics[width=3.4in]{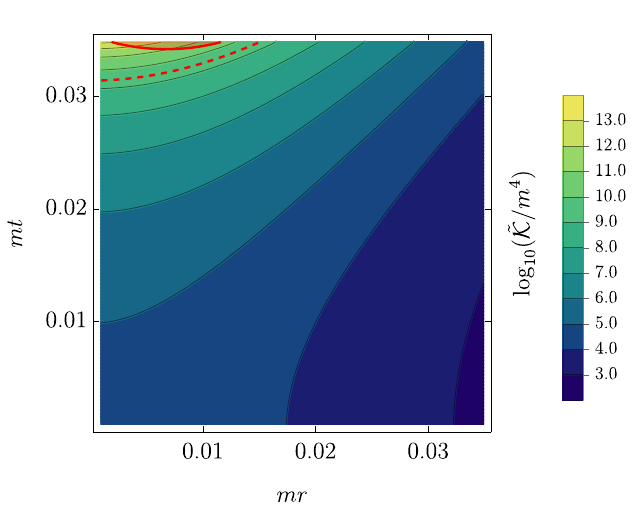}
\caption{The Kretschmann scalar, $\tilde{\mathcal{K}}$, for the $\mu=10^{-2}$ case from Figs. \ref{fig:Euclphixi} and \ref{fig:LiCoInterphi}. The red dashed line is the hyperbolic slice on which $\Theta_\infty=0$, reflecting the formation of the first and outermost trapped surface (at $\tilde{\chi}\rightarrow\infty$). This trapped surface plays the role of the apparent horizon. The red solid line is for $\Theta(\tilde{\xi}(t,r),\tilde{\chi}(t,r))=0$, tracing the formation of the innermost trapped surfaces on subsequent hyperbolic slices. The emerging red shaded area above the red solid line contains trapped surfaces.}
    \label{fig:Kret}
\end{figure}

For a minimally coupled massive scalar field theory with a self interaction $\propto \varphi^n$, the action takes the form,
\begin{eqnarray}
S&=&\int d^4x\sqrt{-g}\Biggl[\dfrac{m_{pl}^2R}{2}
\nonumber\\
&&
-\frac{1}{2}(\partial\varphi)^2-\frac{1}{2}m^2\varphi^2\pm\frac{1}{n}m^2M^2\left(\dfrac{\varphi}{M}\right)^n\Biggr]\,.
\end{eqnarray}
We can rescale the scalar field and the spacetime coordinates in the dimensionless form,
\Beq
\tilde{\varphi}=\frac{\varphi}{M}\,,\qquad \tilde{x}^\mu=mx^\mu\,,
\Eeq
to get
\Beq
S&=\frac{M^2}{m^2}\tilde{S}\,,\\
\tilde{S}&=\int d^4\tilde{x}\sqrt{-g}\left[\frac{\tilde{R}}{2\mu^2}-\frac{1}{2}(\tilde{\partial}\tilde{\varphi})^2-\frac{1}{2}\tilde{\varphi}^2-\frac{1}{n}\tilde{\varphi}^n\right]\,,
\Eeq
where $\mu={M}/{m_{pl}}$.

We see that $\tilde{S}$ depends only on $\mu$,
\Beq
\tilde{S}=\tilde{S}(\mu)\,.
\Eeq
Thus it follows that the Euler-Lagrange equations also yield solutions independent of $M^2/m^2$.

However, for the quantum decay rate, $\Gamma\propto e^{-S_E}$, we do have the dependence on the parameter $M^2/m^2$,
\Beq
\Gamma=\Gamma(M^2/m^2,\mu)\propto \exp\left(-\frac{M^2}{m^2}\tilde{S}_E(\mu)\right)\,,
\Eeq
where $\tilde{S}_E$ is evaluated on the Euclidean solutions of the Euler-Lagrange equations.

\section{$3+1$ ADM formalism}
\label{App:ADM}

We outline our approach to the numerical relativity problem considered in Appendix \ref{App:bubble3plus1}. To this end we adopt the standard 3+1 ADM formalism to describe the variation of the metric, $g_{\mu\nu}$, of the spacetime manifold, $\mathcal{M}$, obeying the Einstein equation

\Beq
G_{\mu\nu}=R_{\mu\nu}-\frac{R}{2}g_{\mu\nu}=\frac{T_{\mu\nu}}{m_{pl}^2}\,.
\Eeq

We foliate $\mathcal{M}$ into three dimensional spacelike hypersurfaces, $\Sigma$, each endowed with a unit normal vector, $n^\mu$, where $n_\mu n^\mu=-1$. It is used to define the induced metric on $\Sigma$, $\gamma_{\mu\nu}=g_{\mu\nu}+n_\mu n_\nu$, which acts as a projection operator, since $\gamma_{\mu\nu}n^\nu=0$. The extrinsic curvature tensor, $K_{\mu\nu}=-\gamma_{\mu}{}^{\alpha}\gamma_{\nu}{}^{\beta}\nabla_{\alpha} n_{\beta}$, determines the variation of the unit normal, as we move along $\Sigma$, and carries information about its embedding in $\mathcal{M}$. Its trace is given by $K=\gamma^{\mu\nu}K_{\mu\nu}$, since it does not vanish only in $\Sigma$. The intrinsic curvature of $\Sigma$ is quantified by its three-dimensional Riemann tensor, defined as ${}^{(3)}R^{\mu}{}_{\nu\alpha\beta}v_\mu=(D_\beta D_\alpha - D_\alpha D_\beta)v_\nu$, where $n_\mu v^\mu=0$, and the three dimensional projection of the covariant derivative is $D_\alpha v^\mu{}=\gamma_\alpha{}^\beta \gamma_\nu {}^\mu\nabla_\beta v^\nu$. The corresponding 3-Ricci tensor is ${}^{(3)}R_{\alpha\beta}={}^{(3)}R^{\mu}{}_{\alpha\mu\beta}$, etc. The $00$ and $0i$ components of the Einstein equation then reduce to the Hamiltonian,
\Beq
{}^{(3)}R+K^2-K_{ij}K^{ij}=\frac{2}{m_{pl}^2}\rho\,,
\Eeq
and momentum
\Beq
D_j(K^{ij}-\gamma^{ij}K)=\frac{\mathcal{S}^i}{m_{pl}^2}\,,
\Eeq
constraints, respectively (also known as the Gauss and Codazzi equations, respectively). Here, the covariant 3-derivative of the 3-tensor is $D_\alpha K^\mu{}_\nu=\gamma_\alpha{}^\beta \gamma_\sigma {}^\mu\gamma_\nu {}^\eta \nabla_\beta K^\sigma {}_\eta$, and the energy and momentum densities are $\rho=n_{\mu}n_{\nu}T^{\mu\nu}$ and $\mathcal{S}^i=-\gamma^{ij}n^{\mu}T_{\mu j}$, respectively, as measured by a normal observer, with a 4-velocity given by the unit norm $n^\mu$.

The unit norm can be expressed in terms of the lapse (scalar) function, $\alpha$, and the shift spatial (living in $\Sigma$) vector, $\beta^i$, as $n^{\mu}=(\alpha^{-1},-\alpha^{-1}\beta^i)$. These determine the choice of coordinate system, allowing to express the invariant line element $ds^2=g_{\mu\nu}dx^\mu dx^\nu$ on $\mathcal{M}$ as
\Beq
ds^2=-\alpha^2dt^2+\gamma_{ij}(dx^i+\beta^i dt)(dx^j+\beta^j dt)\,.
\Eeq
The $ij$ component of the Einstein equation then becomes the first order in time evolution equation for the extrinsic curvature
\Beq
\label{eq:Riccieqn}
\partial_t K_{ij}=&\alpha({}^{(3)}R_{ij}-2K_{ik}K^k{}_j+KK_{ij})-D_iD_j\alpha\\
&-\frac{\alpha}{m_{pl}^2}(\mathcal{S}_{ij}-2^{-1}\gamma_{ij}(\mathcal{S}-\rho))\\
&+\beta^k\partial_k K_{ij}+K_{ik}\partial_j \beta^k+K_{kj}\partial_i\beta^k\,,
\Eeq
also known as the Ricci equation. Here the spatial stress tensor and its trace are $\mathcal{S}^{ij}=\gamma^{ik}\gamma^{il}T_{kl}$ and $\mathcal{S}=\gamma^{ij}\mathcal{S}_{ij}$, respectively. There is also a first order in time evolution equation for $\gamma_{ij}$, coming from the definition of $K_{ij}$, which takes the form
\Beq
\label{eq:Kdef}
\partial_t\gamma_{ij}=-2\alpha K_{ij}+D_i\beta_j+D_j\beta_i\,.
\Eeq
For more details, see, e.g., Ref. \cite{Baumgarte:2010ndz}.

\subsection{Maximal slicing}

To make further progress, we need to pick a coordinate system (i.e., fix the gauge ambiguity due to diffeomorphism invariance) which amounts to choosing the slicing and the threading. We work in maximal slicing in which
\Beq
\label{eq:trK}
K=K^i{}_i=0\,,
\Eeq
in the coordinate system given in eq. \eqref{eq:metric}, assuming spherical symmetry. 

The reason for working in this coordinate system is two fold. Firstly, in anticipation of singularity formation, we choose a singularity avoiding parametrisation of the spacetime metric -- maximal slicing -- since we do not wish to capture the actual singularity and incur uncontrolled numerical errors (even if excision is implemented). Secondly, we picked our coordinate system for its property to capture apparent horizons, which is key for the determination of the nature of the singularity -- whether it is naked or not. Other popular choices, like polar slicing \cite{PhysRevLett.70.9}, do not admit the formation of trapped surfaces (and thus apparent horizons) \cite{Baumgarte:2010ndz} and are unsuitable for our purposes. 

We now derive the closed system of evolution equations for the variables, $\varphi$, $\Pi$, $\Phi$, $\beta$, $\alpha$, $a$, $Q$, where $Q$ is the shear in the radial direction
\Beq
\label{eq:defQ}
Q=\frac{K}{3}-K^r{}_r\,,
\Eeq
$\Pi$ is the conjugate momentum
\Beq
\Pi=-\frac{1}{\alpha}(\partial_t\varphi-\beta\partial_r\varphi)=-\partial_{\perp}\varphi\,,
\Eeq
and $a$ is the radial acceleration, to be defined below.

\begin{figure}[t]
\includegraphics[height=3.4in]{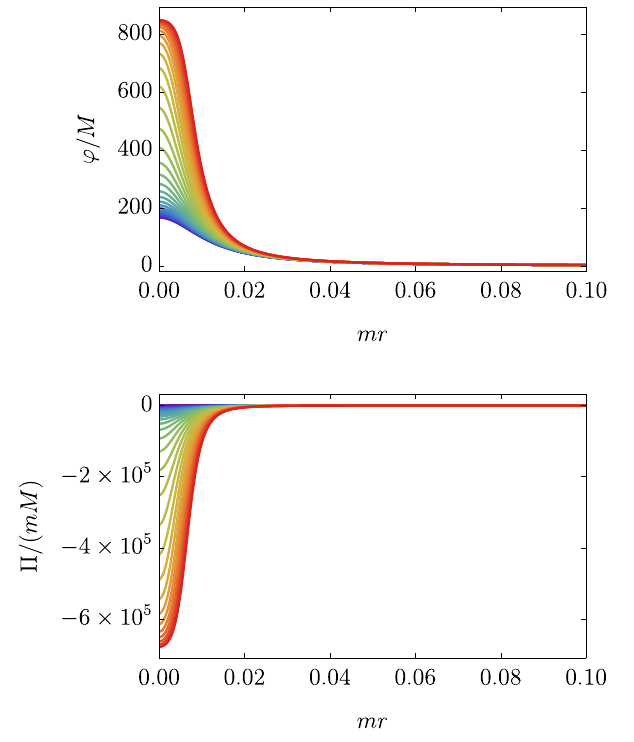}
\caption{Evolution of scalar field (top) and its conjugate momentum (bottom) during the collapse, for $\mu=10^{-3}$ Coleman-De Luccia bubble. Once the scalar field reaches the Planck scale, $\varphi/M\sim1/\mu$, gravitational backreaction becomes important and slows down the evolution. The curves in the figure are plotted at different times, starting from $t=0$ (purple) until $t=0.017168m^{-1}$ (red) with time increment $\Delta t=0.000592m^{-1}$.}
    \label{fig:PhiPi}
\end{figure}
\begin{figure}[t]
\includegraphics[height=3.4in]{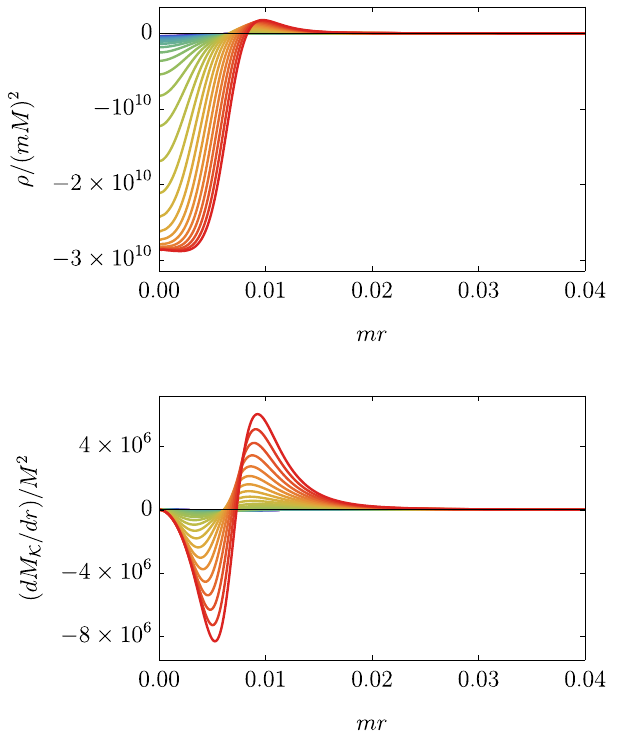}
\caption{Evolution of the energy density $\rho=n^\mu n^\nu T_{\mu\nu}$ (top), and the Kodama mass integrand from eq. \eqref{eq:dMKdr} (bottom), for the bubble from Fig. \ref{fig:PhiPi}. The energy density remains negative in the core region, implying we are capturing the AdS phase in the bubble evolution. The total Kodama mass vanishes. The sustained negativity of the core energy density is reflected in the Kodama mass integrand which starts out negative and only at large distances becomes positive. Since $M_{\mathcal{K}}(r)=\int_0^r (dM_{\mathcal{K}}/d\tilde{r})d\tilde{r}\leq 0$, there is no possibility for formation of a trapped surface for the duration of the $3+1$ ADM simulations.}
    \label{fig:rhodMKdr}
\end{figure}

The closed system of equations is the following:
\Beq
\label{eq:Pired}
\partial_t\varphi=-\alpha\Pi+\beta\partial_r\varphi\,,
\Eeq

\Beq
\label{eq:KGv2}
\partial_t\Pi=\beta\partial_r\Pi-\alpha\left[\partial_r^2\varphi+(a+2\partial_r\ln\Phi)\partial_r\varphi\right]+\alpha\frac{dV}{d\varphi}\,,
\Eeq

\Beq
\label{eq:dtPhi}
\partial_t\Phi=-\alpha\frac{Q}{2}\Phi+\beta\partial_r\Phi\,,
\Eeq

\Beq
\label{eq:beta}
\partial_r\beta=-\alpha Q\,,
\Eeq

\Beq
\label{eq:acc}
\partial_r\ln\alpha=a\,,
\Eeq

\Beq
\label{eq:aeom}
\partial_r a=-a^2-2a\partial_r\ln\Phi+\frac{3}{2}Q^2+\frac{1}{m_{pl}^2}\left[\Pi^2-V\right]\,,
\Eeq

\Beq
\label{eq:momconstr}
\partial_r Q=-3Q\partial_r\ln\Phi -\frac{\Pi}{m_{pl}^2}\partial_r\varphi\,,
\Eeq
where eq. \eqref{eq:Pired} is a re-casted form of the definition of the scalar field conjugate momentum, eq. \eqref{eq:KGv2} is the Klein-Gordan equation, eqs. \eqref{eq:dtPhi} and \eqref{eq:beta} follow from eqs. (\ref{eq:Kdef}) and (\ref{eq:defQ}), eq. \eqref{eq:acc} is the definition of the acceleration, eq. \eqref{eq:aeom} is the contracted version of the Ricci equation, eq. \eqref{eq:Riccieqn}, in maximal slicing, and eq. \eqref{eq:momconstr} is the momentum constraint.

In addition the Hamiltonian constraint reduces to

\Beq
\frac{\partial_r^2\Phi}{\Phi}=&\frac{1-(\partial_r\Phi)^2}{2\Phi^2}\\&-\frac{3}{8}Q^2-\frac{1}{2m_{pl}^2}\left[\frac{\Pi^2}{2}+\frac{(\partial_r\varphi)^2}{2}+V\right]\,.
\Eeq

We solve the system of eqs. (\ref{eq:Pired}-\ref{eq:momconstr}) using a forward finite difference discretisation of the spacetime derivatives of the metric, and of the spatial derivatives of the scalar field, whereas for the time derivatives of the scalar field we use the staggered leapfrog method. We use uniform spatial and temporal grids, with grid spacings $\delta r=10^{-4}m^{-1}$, and $\delta t=10^{-6}m^{-1}$, respectively. In the radial direction we used between $1500$ and $6000$ grid points to verify that our results are robust to changes in the grid size. Reductions in $\delta r$ and/or $\delta t$ by factors $2$ also do not affect the results.

As spatial boundary conditions we use at $r=0$: $\partial_r\varphi=0$, $\partial_r \Phi =1$, $\beta =0$, $a=0$, $ Q=0$, and at the edge of the simulation box, $r=L$: $\alpha =1$.

%\subsubsection{Kretschmann scalar}

%As a measure of the spacetime curvature, we consider the Kretschmann scalar, which is defined in terms of the Riemann tensor as
%\Beq
%\tilde{\mathcal{K}}\equiv R^{\mu\nu\alpha\beta}R_{\mu\nu\alpha\beta}\,.
%\Eeq
%The explicit expression for $\tilde{\mathcal{K}}$ for the metric from eq. \eqref{eq:metric} is cumbersome. However, if one uses the Einstein equations, the Kretschmann scalar simplifies to 
%\Beq
%\label{eq:Kret}
%\tilde{\mathcal{K}}=&\frac{1}{m_{pl}^2 \Phi ^2}\Big[ \left(4+(\Phi Q)^2-4(\partial_r\Phi)^2\right)\left((\partial_r\varphi)^2-\Pi ^2-2 V\right)\\
%&-24  (\partial_r\Phi-1) \Pi 
%   Q \Phi\partial_r\varphi\Big]\\
%   &+\frac{2}{m_{pl}^4}
%   \left[\left((\partial_r\varphi)^2-\Pi ^2\right)^2+2 V^2\right]+\frac{12}{\Phi ^4}\left[(\partial_r\Phi)^2-1\right]^2\\
%   &-\frac{6}{\Phi ^2}(\partial_r\Phi-1) (7 \partial_r\Phi -5) Q^2+\frac{3}{4}Q^4\,.
%\Eeq

\begin{figure}[t]
\includegraphics[height=3.4in]{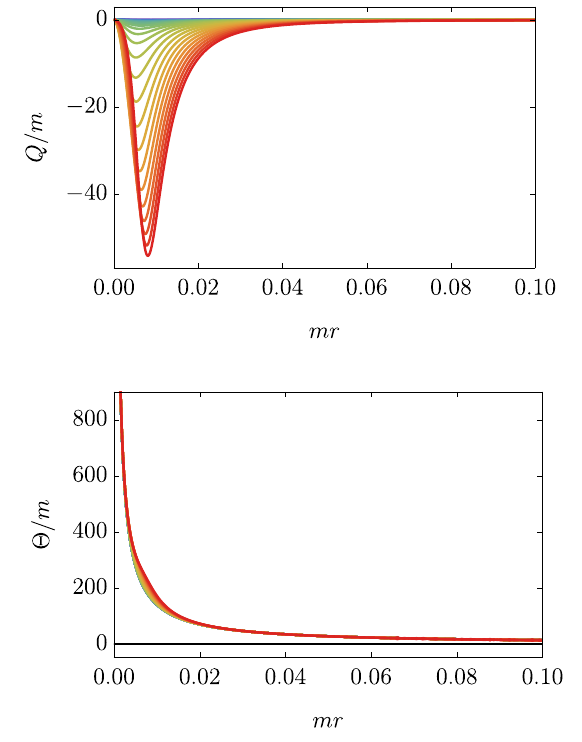}
\caption{The radial shear (top) and expansion rate of an outgoing null congruence (bottom) for the bubble from Fig. \ref{fig:PhiPi}. Since in the maximal slicing we work in $Q=-K^r_r$, the negative shear implies contraction in the radial direction, which is a manifestation of the collapse of the bubble interior. $\Theta>0$ throughout, which can be understood as a consequence of the non-positive Kodama mass, and implies that no trapped surfaces are formed for the duration of the $3+1$ ADM simulations.}
    \label{fig:QTheta}
\end{figure}
\begin{figure}[t]
\includegraphics[height=3.4in]{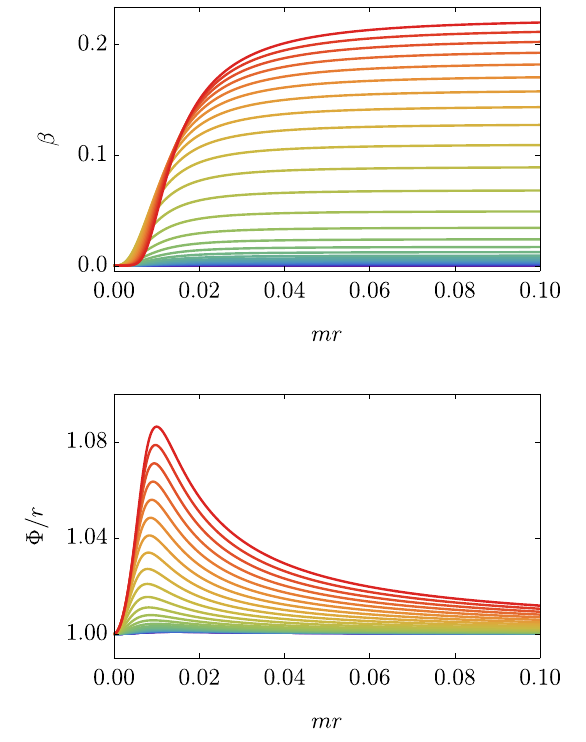}
\caption{The radial shift (top) and the areal radius (bottom) for the bubble from Fig. \ref{fig:PhiPi}. During the collapse, the spacetime threading becomes increasingly sparse outside the bubble, which is reflected by the growth of $\beta$ with time on large distances (and sometimes referred to as `coordinate stretching' \cite{Baumgarte:2010ndz}). The AdS bubble is roughly uniform, giving rise to an approximately linearly growing with distance $\Phi/r$ within the core. Outside the bubble, $\Phi/r$ decays inversely with distance and eventually settles to an $r$-independent value, since the bubble has a vanishing Kodama (or equivalently ADM) mass. The value grows with time (reflecting the stretching of radial threading).}
    \label{fig:betaPhi}
\end{figure}

\section{Bubble evolution in $3+1$ ADM}
\label{App:bubble3plus1}

The gravitational bubble evolution is studied for $t\geq0$ in the spacetime metric,
\Beq
\label{eq:metric}
ds^2=&-\alpha^2dt^2+(dr+\beta dt)^2+\Phi^2d\Omega_2^2\,,\\
d\Omega_2^2=&d\theta^2+\sin^2\theta d\phi^2\,.
\Eeq
The evolution of the lapse, $\alpha(t,r)$, radial shift, $\beta(t,r)$, areal radius, $\Phi(t,r)$, and the scalar field is calculated in the $3+1$ formalism in the maximal slicing gauge in which the mean extrinsic curvature vanishes, $K=0$. Its only non-vanishing component is given by the radial shear $Q(t,r)\equiv(K/3)-K^r_r=K^\theta_\theta/2=K^\phi_\phi/2$. For more details, see Appendix \ref{App:ADM}.

\begin{figure}[t]
\par\bigskip
\vspace{-0.3cm}
\includegraphics[trim={0 0 0 0.075cm}, clip, height=1.63in]{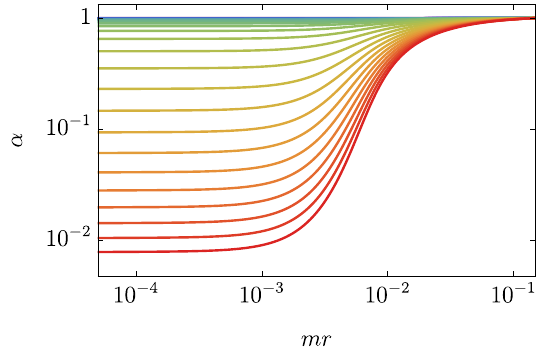}
\caption{The lapse function during the collapse of the bubble from Fig. \ref{fig:PhiPi}. As the bubble becomes increasingly curved, we observe an exponential decrease of $\alpha$ within its interior (`the collapse of the lapse' \cite{Baumgarte:2010ndz}), which is a consequence of the singularity avoiding property of the maximal slicing.}
    \label{fig:alpha}
\end{figure}

The initial data for the scalar field and the metric at $t=0$ is given by the Coleman-De Luccia solution from Section \ref{sec:Bubnuc} after making the identification $\xi\rightarrow \sqrt{r'^2-t^2}=r'$, and $r=\int_0^{r'(r)}d\tilde{r}'\omega(\tilde{r}')$, as well as recalling that the bounce is time-reversal symmetric (and nucleation happens at its half-way point, so time derivatives should vanish). In particular, 
\Beq
\varphi(t=0,r)&=\varphi(\xi\rightarrow r'(r))\,,\\
\partial_t\varphi(t=0,r)&=0\,,\\
\alpha(t=0,r)&=\omega(\xi\rightarrow r'(r))\,,\\
\Phi(t=0,r)&=r'(r)\omega(\xi\rightarrow r'(r))\,,\\
\beta(t=0,r)&=0\,,\\
Q(t=0,r)&=0\,.
\Eeq
The initial data satisfies the Hamiltonian and momentum constraints exactly (by construction). Furthermore, it gives rise to a vanishing ADM mass, which due to spherical symmetry can be given by the Misner-Sharp mass \cite{PhysRev.136.B571},
\Beq
\label{eq:dMKdr}
M_{\rm MS}=\lim_{r\rightarrow\infty}4\pi m_{pl}^2\Phi\left[1+\frac{(\Phi Q)^2}{4}-(\partial_r\Phi)^2\right]\,.
\Eeq
The spherical symmetry also implies the existence of a Killing-like vector, known as the Kodama vector, $\mathcal{K}^\mu$, \cite{Kodama:1979vn}, which in turn gives rise to a conserved current, $T^\mu{}_\nu\mathcal{K}^\nu$, known as the Kodama current. The corresponding conserved charge is called the Kodama mass, and is the $r\rightarrow\infty$ limit of the function $M_{\mathcal{K}}(r)=4\pi\int_0^r d\tilde{r} \alpha\Phi^2T^t{}_\mu\mathcal{K}^\mu$, which also equals the Misner-Sharp mass. For our initial data, the net Kodama mass vanishes, $M_{\mathcal{K}}(r\rightarrow\infty)=0$. For our choice of slicing, the explicit expression for the Kodama mass integrand is
\Beq
\frac{dM_{\mathcal{K}}}{dr}=4\pi\Phi^2\left[ \left(\frac{\Pi^2}{2}+\frac{(\partial_r\varphi)^2}{2}+V\right)\partial_r\Phi-\frac{Q\Phi\Pi\partial_r\varphi}{2}\right]\,,
\Eeq
where we used the Kodama vector components $\mathcal{K}^t=-\alpha^{-1}\partial_r\Phi$ and $\mathcal{K}^r=\alpha^{-1}\partial_t\Phi$, and the scalar field conjugate momentum is given by $\Pi=-\alpha^{-1}(\partial_t\varphi-\beta\partial_r\varphi)\equiv-\partial_{\perp}\varphi$.

\begin{figure}[t]
\includegraphics[height=1.7in]{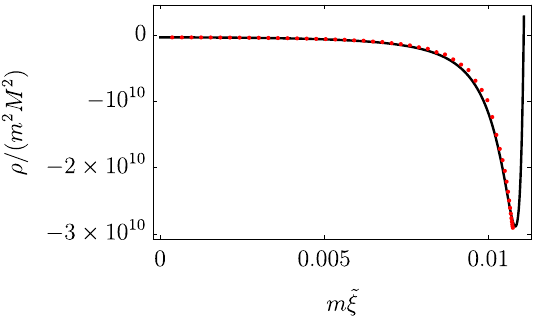}
\caption{The evolution of the energy density for $\mu=10^{-3}$ according to eqs. \eqref{eq:LorLCintKG} and \eqref{eq:LorLCintddomega} for an $O(3,1)$ bubble interior (black solid line), and according to the $3+1$ ADM simulations eqs. (\ref{eq:Pired}-\ref{eq:momconstr}) for a spatially spherically symmetric bubble (red dots), where $\int_0^{\tilde{\xi}(t)} d\tilde{\xi}'\omega(\tilde{\xi}')=\int_0^t dt'\alpha(t',r=0)$ ($t$ and $r$ are from eq. \eqref{eq:metric}). For their duration, the ADM simulations reproduce the $O(3,1)$ solution at $\mathcal{O}(1\%)$. In particular, they capture accurately the initial AdS stage of monotonically decreasing energy density. Numerical instabilities prevent them from resolving the subsequent abrupt (``stiff'') upward turn and the following rapid growth of $\rho$ to positive values, accompanied by trapped surface generation and ultimate crunch singularity formation. Nevertheless, the agreement of the two methods within the overlap $\tilde{\xi}$ range serves as a proof that the assumption of an $O(3,1)$-symmetric bubble interior is valid. Note that if the $O(3,1)$ symmetry applies for a finite range of $\tilde{\xi}$, then it holds throughout the bubble interior by virtue of the evolution eqs. \eqref{eq:LorLCintKG} and \eqref{eq:LorLCintddomega}.}
    \label{fig:rhosim}
\end{figure}

The bubble collapse proceeds as follows. In the core region $\varphi$ rolls down the negative slope (towards $\varphi>M$), see Fig. \ref{fig:PhiPi}. The evolution slows down as $\varphi\sim M/\mu=m_{pl}$. This is the moment when the growing compaction function 
$\mathcal{C}$~\cite{Shibata:1999zs} of the core region becomes $\mathcal{O}(1)$, gravity becomes strong, and gravitational backreaction sizeable.\footnote{Roughly, $\mathcal{C}\sim (m_{pl}^2R)^{-1}\int^R_0 dr |\rho| r^2\sim (m_{pl}^2R)^{-1}R^3(\varphi/R)^2\sim (\varphi/m_{pl})^2$, and thus is of order unity for Planck-scale field values, regardless of the size of the core region, $R$, provided the bubble wall is not thin, which is applicable in our case.}

In addition to the evolution of the field, on our time slices $\Pi$ itself speeds up. This implies that the kinetic energy of the scalar field grows in comparison with the magnitude of the potential energy, $V$. Since the field is deep in the negative potential region, the total energy density of the scalar field, $\rho$, slows down the increase of its negativity. 
See the top panel in Fig. \ref{fig:rhodMKdr}. 
This is also reflected in the Kodama mass (see bottom panel in Fig. \ref{fig:rhodMKdr}), whose integrand is negative within the core region, but on larger distances turns positive, since the total $M_{\mathcal{K}}(r\rightarrow\infty)=0$. 

Another manifestation of the onset of gravitational backreaction is the increase in the extrinsic curvature, as shown in the top panel in Fig. \ref{fig:QTheta}, where $Q=-K^r_r$ becomes substantial in accordance with $\mathcal{C}\sim1$. 
More specifically, $Q$ becomes increasingly negative (indicating contraction in the radial direction) and at late times in the core region $Q^2\sim -m_{pl}^{-2}\rho\sim (m/\mu)^2$.\footnote{The second relation comes from $\Pi^2/2\sim -V$ with $\rho\sim\Pi^2/2+V\sim V<0$.} Given that we are approaching a spacetime region of large (extrinsic) curvature, we check if any trapped surfaces (and consequently any apparent horizons) form. To this end we implement the standard procedure from Section \ref{eq:Ahf} (with $s^\mu=(0,1,0,0)$). In particular, we calculate the expansion of the outgoing null congruence, given by
\Beq
\Theta = \frac{\sqrt{2}}{\Phi}\left(\partial_{\perp}\Phi+\partial_r\Phi\right)\,.
\Eeq
As before, if $\Theta$ ever becomes negative, we have a trapped surface, and the outermost such surface is the apparent horizon. As shown in the bottom panel in Fig. \ref{fig:QTheta}, $\Theta$ remains positive throughout, and no apparent horizons form in our ADM $3+1$ simulations. 

\begin{figure}[t]
\includegraphics[trim={0 1cm 0 3cm},clip]{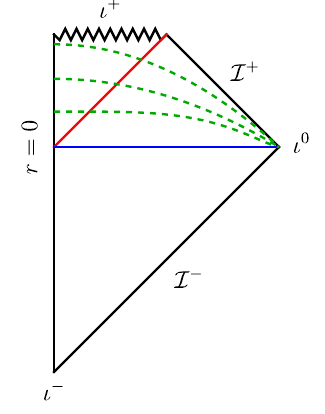}
\caption{An illustration of the spacetime of an AdS bubble, embedded in asymptotically flat spacetime, undergoing a gravitational collapse. The blue line depicts the nucleation -- the tunneling process giving rise to the formation of a bubble, which in our work we assumed to be given by a Coleman-De Luccia bounce. The subsequent collapse of the bubble leads to the formation of a space-like singularity, $\mathcal{\iota}^+$ -- a crunch. The collapse occurs only within the interior of the bubble, whereas the bubble expands in the false-vacuum exterior, eventually devouring the entire spacetime. The null red geodesic plays the role of a cosmological horizon. The dashed green lines depict the maximal slices ($K=0$) used in our $3+1$ ADM numerical studies. Unlike cases with bounded negative $V$, e.g.,  \cite{Abbott:1985kr}, in which the crunch singularity forms as the scalar field oscillates about the negative true vacuum, our unbounded potential gives rise to trapped surfaces and subsequent crunch formation, while the scalar field is rolling down the slope in the negative $V$ region. Our discovery shows that bounding $V$ from below is not a necessary condition for singularity formation.}
\label{fig:Penrose}
\end{figure}

In Fig. \ref{fig:betaPhi} we show the evolution of the radial shift, $\beta$, and areal radius $\Phi$. The growth with time of $\beta$ on large distances reveals that fewer spatial threads remain on large $r$. This coordinate stretching is typical of singularity avoiding slicing choices like maximal slicing. A similar pattern is observed for $\Phi/r$ at large $r$ (i.e., a nearly $r$-independent value growing with time due to coordinate stretching), whereas for smaller $r$ it exhibits the standard behaviour of a gravitational potential of a spherical object -- a linear growth with distance within the core, and then a decay inversely proportional to $r$. The deteriorating resolution on large $r$ was a challenge to our simulations, and we made sure that the violation of the Hamiltonian constraint, as well as the conservation of the net Kodama mass were $\mathcal{O}(1\%)$. We also ensured that using a simulation box of fixed radial coordinate extend, did not lead to a non-negligible violation of energy conservation\footnote{Recall that $\nabla_\mu T^{\mu\nu}=\partial_\mu(\sqrt{-g}T^{\mu\nu})+\sqrt{-g}\Gamma_{\alpha\beta}^\nu T^{\alpha\beta}=0$. For a spacetime volume of fixed spatial coordinate size (e.g., $r$), the spacetime integral of $\nabla_\mu T^{\mu\nu}$ may receive contributions from the spatial boundary if the term $\sqrt{-g}\Gamma_{\alpha\beta}^\nu T^{\alpha\beta}$ does not decay with distance sufficiently fast. For the finite extend of the box to not lead to significant violation of energy conservation, we need the time-slice warping $\Gamma_{ij}^0\propto Q$ to decrease at least as fast as $r^{-3}$ for a non-growing $T^{\mu\nu}$ with $r$.}, by verifying that $Q\propto r^{-3}$ on large $r$.

The slowing down of the evolution of the system is a consequence of the exponential decay of the lapse in the core region, $\ln\alpha\propto -t$, depicted in Fig. \ref{fig:alpha}, making a coordinate time-interval correspond to an exponentially small proper time interval, $d\tau'=\sqrt{\alpha^2-\beta^2} dt\approx\alpha dt$, of a static observer in the core. This behaviour of $\alpha$ is typical for singularity avoiding slicing choices, like the one we employ -- maximal slicing. In particular, the `collapse' of $\alpha$ slows down the progression of the time slices in regions of high curvature. This is indeed the case, as shown in Fig. \ref{fig:rhosim}, where we see that the substantial slow-down of $\alpha$ reflects the approach to a point of a rapid change in $\rho$ -- from decreasing to increasing, which sources huge spacetime curvature. Our $3+1$ ADM simulations cannot evolve the system beyond this point -- they fail to resolve the rapid changes, and numerical instabilities accompanied by large violations of constraints, render subsequent results unreliable. None the less, the simulations reproduce accurately, at $\mathcal{O}(1\%)$ the $O(3,1)$ solution for the bubble interior, proving that the crucial simplifying assumption of an $O(3,1)$ symmetric interior is legitimate.

Finally, in Fig. \ref{fig:Penrose} we summaries the spacetime picture confirmed by our $3+1$ ADM simulations.

\bibliography{ref}

\end{document}